\documentclass[12pt,tightenlines,eqsecnum,floats,showpacs,nofootinbib,amsmath,amssymb,aps,prd,superscriptaddress]{revtex4-1}

\usepackage{graphicx}
\usepackage{amsmath,amssymb}
\usepackage{hyperref}
\usepackage{slashed}

\begin{document}
	
\title{Uniqueness of the Fock quantization of Dirac fields in 2+1 dimensions}
	
\author{Jer\'onimo Cortez}
\email{jacq@ciencias.unam.mx}
\affiliation{Departamento de F\'isica, Facultad de Ciencias, Universidad Nacional Aut\'onoma de M\'exico, M\'exico D.F. 04510, M\'exico}
\author{Beatriz Elizaga Navascu\'es}
\email{beatriz.elizaga@iem.cfmac.csic.es}
\affiliation{Instituto de Estructura de la Materia, IEM-CSIC, Serrano 121, 28006 Madrid, Spain}
\author{Mercedes Mart\'in-Benito}
\email{m.martin@hef.ru.nl}
\affiliation{Radboud University Nijmegen, Institute for Mathematics, Astrophysics and Particle Physics, Heyendaalseweg 135, NL-6525 AJ Nijmegen, Netherlands}
\affiliation{Instituto de Astrof\'isica e Ci\^encias do Espa\~co, Universidade de Lisboa, Faculdade de Ci\^encias, Ed. C8, Campo Grande, PT1749-016 Lisboa, Portugal}
\author{Guillermo A. Mena Marug\'an} \email{mena@iem.cfmac.csic.es}
\affiliation{Instituto de Estructura de la Materia, IEM-CSIC, Serrano 121, 28006 Madrid, Spain}
\author{Jos\'e M. Velhinho}
\email{jvelhi@ubi.pt}
\affiliation{Universidade da Beira Interior, Rua Marqu\^es d'\'Avila e Bolama, 6201-001, Covilh\~a, Portugal}

\begin{abstract}
We study the Fock quantization of a free Dirac field in 2+1-dimensional backgrounds which are conformally ultrastatic, with a time-dependent conformal factor. As it is typical for field theories, there is an infinite ambiguity in the Fock representation of the canonical anticommutation relations. Different choices may lead to unitarily inequivalent theories that describe different physics. To remove this ambiguity one usually requires that the vacuum be invariant under the unitary transformations that implement the symmetries of the equations of motion. However, in nonstationary backgrounds, where time translation is not a symmetry transformation, the requirement of vacuum invariance is in general not enough to fix completely the Fock representation. We show that this problem is overcome in the considered scenario by demanding, in addition, a unitarily implementable nontrivial quantum dynamics. The combined imposition of these conditions selects a unique family of equivalent Fock representations. Moreover, one also obtains an essentially unique splitting of the time variation of the Dirac field into an explicit dependence on the background scale factor and a quantum evolution of the corresponding creation and annihilation operators. 

\end{abstract}

	
\maketitle

\section{Introduction}
\label{sec:Intro}

The quantization of a classical system involves ambiguities. Different choices in the process may lead to different quantum theories, and these may turn out to correspond to different physics. This is a serious drawback if one wants to arrive to robust quantum predictions. Focusing on the canonical quantization of linear theories, already at the classical level we have the freedom of choosing different sets of canonical variables to describe the system. At the quantum level, there is also ambiguity in adopting a representation for their corresponding canonical commutation or anticommutation relations. In standard quantum mechanics, to remove these ambiguities one can appeal to the uniqueness Stone-von Neumann theorem. This theorem guarantees that all possible representations of the Weyl algebra that verify strong continuity, unitarity, and irreducibility are unitarily equivalent \cite{svn}. In quantum field theory (QFT) the situation is more complicated. Even if we restrict the attention to Fock representations of the field analogue of the Weyl algebra, there is no counterpart of the Stone-von Neumann theorem owing to the infinite number of degrees of freedom  \cite{wald}. The usual way to proceed is to make use of the symmetries of the system.  For Fock representations, one usually requires the invariance of the vacuum under the symmetry transformations, leading to a natural implementation of those transformations  as unitary operators.  This strategy eliminates the ambiguity in the case of QFT in highly  symmetric backgrounds, the prototypical  example being QFT in flat spacetime, where Poincar\'e invariance fixes a unique Minkowski vacuum \cite{wald}. Also in the case of stationary backgrounds, where time translations form a group of isometries, invariance under these transformations and positiveness of the energy are sufficient to remove the ambiguity \cite{kay, baez}.  Nevertheless, when the field evolves in a nonstationary background, the requirement of invariance of the vacuum under  spatial symmetry transformations might not be enough to pick out a unique Fock representation, and the question arises whether one can appeal to extra conditions in order to remove the remaining ambiguity \cite{ashmag,jac}. 

This question has an affirmative answer for the case of a real Klein-Gordon (KG) field minimally coupled to geometries with homogeneous and isotropic compact spatial sections of dimension smaller than four \cite{compact,compact2}, which in particular are relevant in cosmology \cite{unit}. In fact, in such scenarios, the absent time-translation invariance is replaced with a demand of unitary implementability of the dynamics in Fock space. This requirement, together with the invariance of the vacuum under the group of spatial symmetries, singles out a unique explicitly time-dependent parametrization of the scalar field and a unique class of unitarily equivalent Fock representations for the associated canonical commutation relations. The privileged choice of canonical variables involves a scaling of the original KG field by a homogeneous function of the background, such that the rescaled field behaves as a KG field with time-dependent mass propagating in an ultrastatic spacetime. Actually, the rescaled field can be seen as conformally coupled to the original geometry (but massive). Furthermore, the remaining freedom in the choice of a canonical momentum for the rescaled field is removed as well \cite{uniqmom}. On the other hand, the condition of unitary implementability of the dynamics translates into a regularity condition in the ultraviolet regime of large eigenvalues of the Laplace-Beltrami operator defined on the spatial sections\footnote{This has some similarities with the ultraviolet regularity imposed on Hadamard states. For spacetimes with compact spatial sections, the set of Hadamard states provides a well-known unitary equivalence class of quantum representations. For a discussion on the relation between the unitary dynamics approach and Hadamard states see Ref. \cite{compact2}.}. A simple representative in the equivalence class of vacua selected with this criterion is the conformal vacuum that would be naturally selected if the field were massless. Indeed, in the ultraviolet regime, the dynamical behavior of the massive system tends fast enough to that of the massless situation, and conformal invariance is approximately recovered.

Beyond the case of the KG field, these uniqueness results have been recently extended to Dirac fields with minimal coupling propagating in a cosmological Friedmann-Lema$\hat{\rm {\i}}$tre-Robertson-Walker background with closed spatial sections \cite{uf}. Again the criterion of invariance of the vacuum under the group of spatial symmetries, together with the unitary implementability of the dynamics, selects a unique family of equivalent Fock representations for the associated canonical anticommutation relations. The unitary implementability of the evolution now imposes regularity conditions in the ultraviolet regime of large eigenvalues (in absolute value) of the Dirac operator defined on the spatial sections\footnote{Just like in the scalar field case, there is the possibility that suitably defined fermionic states with Hadamard properties may provide a natural unitary equivalence class of quantum
representations. To the best of our knowledge this remains an open issue, which falls outside
the scope of the present work. We present here a different approach, insisting on quantum unitary dynamics, which we believe is a desirable aspect in quantum physics.}. Such a uniqueness result is established under very mild conditions on the time variation of the cosmological background, and once a convention on the notions of particle and antiparticle has been set. Interestingly, part of the necessary change of parametrization in the description of the fermion field is again the same homogeneous scaling that would correspond to a conformally coupled field, and appears here to render constant canonical anticommutation relations. However, in contrast with the KG field case, this is not the final field parametrization compatible with a unitarily implementable dynamics. This unitarity requires a very specific description of the fermion modes of the field, when we expand it in a basis of eigenspinors of the Dirac operator. More concretely, in the limit of large Dirac eigenvalues, the time-dependent factors of the dominant terms of the particle and antiparticle parts of the field are uniquely fixed up to phases, for the two possible chiralities. These time-dependent factors are different for the particle and antiparticle contributions, and involve the mass of the fermion field and the scale factor of the cosmology under study. Therefore, the demand of a unitary dynamics  introduces a different time-dependent parametrization for the different parts of the Dirac field. This is due to the fact that, unlike to the situation found for the KG field, the Dirac evolution cannot be treated asymptotically as if it were conformally invariant, since the mass term couples the two chiralities of the Dirac field and gives a nonnegligible contribution to the dynamics in the ultraviolet regime. These results provide an example of how it may be possible to implement the dynamics as a unitary endomorphism even if the system does not display an asymptotic conformal invariance in the ultraviolet.

The aim of this paper is to further extend the above uniqueness results to the Fock quantization of both massive and massless fermion fields in 2+1 dimensions. Two main reasons motivate our study. First of all, it is interesting to compare the analysis of fermion fields for different dimensions since, owing to their spinorial nature, they present differences that do not appear in the case  of a scalar field. Secondly, Dirac fields do not only play a role in general relativity, but appear as well in lower dimensions in condensed matter systems. For example, excitations in graphene obey the dynamics of a Dirac fermion in 2+1 dimensions. Usually, these excitations behave as corresponding to massless fermions \cite{graphene}, but in certain situations the energy spectrum also displays a finite gap corresponding to massive fermions \cite{gap}. Therefore, our results might contribute to improve the foundations of the theoretical studies of fermions in this type of materials. 

We will prove that the requirement of invariance of the vacuum under the symmetries of the Dirac equation, together with the requirement of a unitarily implementable nontrivial fermion dynamics, is again a valid criterion to select a unique family of unitarily equivalent Fock representations, up to the convention fixed for the concepts of particles and antiparticles. Furthermore, the criterion determines a quantum evolution for the resulting creation and annihilation operators of the Dirac field, except for unitary redefinitions that are irrelevant in the ultraviolet sector. The remaining time variation of the field is given by its explicit dependence on the background scale factor. 

In comparison with the 3+1 dimensional case, now the field does not present two chiralities, although the fermion dynamics does couple the parts of the field corresponding to Dirac eigenvalues of opposite sign. And this coupling is such that, at first sight, one might tend to believe that the system would behave as a conformally invariant massless field in the ultraviolet regime of large Dirac eigenvalues. Then, based on the situation found for the massive KG field, one would be tempted to conclude that the conformal vacuum ought to lay in the selected family of unitarily equivalent Fock representations for the massive field. Nevertheless, this is not true. As we will see, the mass still gives a nonnegligible contribution to the dynamics in the ultraviolet limit, precisely owing to its presence in the dynamical coupling that we have mentioned above.

The structure of the paper is as follows. In Sec. \ref{sec:Model} we describe the classical model and perform an asymptotic analysis of its dynamics. In Sec. \ref{sec:CS} we introduce the complex structure, mathematical object that characterizes a Fock representation and then codifies the infinite ambiguity in its choice. Besides, we consider the additional ambiguity present in the choice of canonical variables by contemplating the possibility of performing canonical time-dependent transformations. We characterize the set of complex structures that lead to a vacuum that is invariant under the symmetry group of the Dirac equation. Then, in Sec. \ref{sec:Unitarity}, we study under which conditions the complex structure and the choice of fundamental variables for the field description allow for a nontrivial unitary implementation of the dynamics in Fock space. Section \ref{proof} contains the proof of the unitary equivalence between all the admissible Fock quantizations selected by our criterion, provided a convention for particles and antiparticles is given. Finally, in Sec. \ref{sec:con} we summarize and discuss our results.

\section{The classical system}
\label{sec:Model}

We consider a (test) fermion field minimally coupled to a fixed three-dimensional and globally hyperbolic differentiable manifold with the topology of $\mathbb{I}\times\Sigma$, where $\mathbb{I} \subseteq \mathbb{R}$ is a connected real interval, and $\Sigma$ is a connected, compact, and orientable two-dimensional Riemannian manifold. The metric is taken to have the form\footnote{We employ Einstein's summation convention.}
\begin{align}\label{metric}
ds^{2}=g_{\mu\nu}\text{d}x^\mu \text{d}x^\nu= e^{2\alpha(\eta)}\left(-\text{d}\eta^{2}+h_{ij}(\vec{x})\text{d}x^i\text{d}x^j\right).
\end{align}
Greek indices run from 0 to 2, Latin indices  from the middle of the alphabet run from 1 to 2, and $\vec{x}=(x^1,x^2)$. Notice that the studied geometries are conformally stationary. The scale factor $e^{\alpha}$ encodes all the nonstationary information of the metric. Up to this factor, $h_{ij}$ is the metric induced on the spatial surfaces defined at each constant value of the conformal time $\eta$. Dirac fields couple to the geometry through the (inverse of the) frame field $e_\mu^a$, defined such that $g_{\mu\nu}= e_\mu^a e_\nu^b \eta_{ab}$. Here, $\eta_{ab}$ is the Minkowski metric in three dimensions, given by $\text{diag}(-1,1,1)$. Latin indices from the beginning of the alphabet account for the extra gauge degrees of freedom that the frame field introduces. These indices are raised and lowered with $\eta^{ab}$ and its inverse $\eta_{ab}$, respectively.

According to our previous comments, the frame field is defined up to $SO(2,1)$ (orthochronous) gauge transformations. Besides, such frame can be defined globally, as the considered three-dimensional manifold is space-time orientable \cite{Stiefel}. Therefore, it admits a spin structure and we can define spinor fields on it. In three dimensions, complex fermion fields are locally represented by two-component spinors $\Psi$, since any of the two irreducible complex representations of the corresponding Clifford algebra are generated by $2\times 2$ Dirac matrices $\gamma^a$, that satisfy $\gamma^{a}\gamma^{b}+\gamma^{b}\gamma^{a}=2\eta^{ab}I_{2\times2}$, where $I_{2\times2}$ is the identity matrix \cite{SGeom}. For concreteness, we will take
\begin{align}
\gamma^0=i\begin{pmatrix}
-1 & 0 \\ 0 & 1
\end{pmatrix},\quad \gamma^1=i\begin{pmatrix}
0 & -1 \\ 1 & 0
\end{pmatrix}, \quad \gamma^2=\begin{pmatrix}
0 & 1 \\ 1 & 0
\end{pmatrix}.
\end{align}
Besides, the two components of the Dirac spinor will be taken to be Grassmann variables, in order to capture the anticommuting nature of the fermion field \cite{Berezin}. For a fermion field of mass $m$, the action of the system is
\begin{align}\label{action}
I_f=-i\int \text{d}^3x \sqrt{-\text{det}(g_{\mu\nu})}\left[\frac12(\Psi^{+} e^\mu_a \gamma^a \nabla^S_\mu \Psi-\text{H.c.})+m\Psi^{+}\Psi\right].
\end{align}
Here, H.c. stands for Hermitian conjugate. We have introduced the notation $\Psi^{+}=\Psi^\dagger \gamma^0$ for the Dirac adjoint, where the dagger denotes the Hermitian adjoint, and $\nabla^S_\mu$ denotes the spin lifting of the Levi-Civita covariant derivative \cite{SGeom}. Its action on the spinors is locally given by
\begin{align}
\nabla^S_\mu \Psi=\partial_\mu\Psi-\frac14 \omega_\mu^{ab} \gamma_b\gamma_a\Psi,
\end{align}
where $\omega_\mu^{ab}$ provides the spin connection one-form,
\begin{align}
\omega_\mu^{ab}=\frac12\left(e^{\nu a}\partial_\mu e^b_\nu + e^{\nu a}e^{\lambda b}\partial_\lambda g_{\mu\nu}-e^{\nu b}\partial_\mu e^a_\nu -e^{\nu b} e^{\lambda a}\partial_\lambda g_{\mu\nu}\right).
\end{align}
In view of the metric \eqref{metric}, it is convenient to perform a partial gauge fixing by choosing $e^0_j=0$. This amounts to a well-defined reduction of the structure group of the bundle of oriented frames from $SO(2,1)$ (orthochronous) to $SO(2)$ \cite{Isham}. In turn, it induces a natural restriction of the spin structure to the spin bundle that double covers the reduced frame bundle. Such a restriction is well-defined and can be understood to provide a spin structure on each of the two-dimensional spatial surfaces \cite{SGeom}. This gauge fixing will allow us to separate the time dependence of the Dirac spinors from their spatial dependence in our analysis\footnote{Once this gauge reduction has been performed, for each value of the time parameter, the fields behave as spinors geometrically defined on each of the two-dimensional spatial manifolds that foliate the background.}. In particular, we will be able to deal directly with the spectral analysis of the Dirac operator defined on the two-dimensional surfaces of constant time, rather than with the Dirac operator on the whole Lorentzian geometry. Indeed, in this gauge, the covariant derivative acting on the spinors locally simplifies to
 \begin{align}
 \nabla^S_0 \Psi=\partial_0 \Psi,\quad \nabla^S_j \Psi={}^{(2)}\nabla^S_j\Psi-\frac14 \tilde\omega_j^{ab} \gamma_b\gamma_a \Psi,
\end{align}
where ${}^{(2)}\nabla^S_j$ is the spin covariant derivative on the spatial slices with metric $h_{ij}$, which are Riemannian two-manifolds, and 
\begin{align}
\tilde\omega_j^{ab}=\frac12\left(e^{i a}e^{0 b}\partial_0 g_{ij}- e^{i b}e^{0 a}\partial_0 g_{ij}\right).
\end{align}
It is easy to check that then we locally have
\begin{align}
e^\mu_a \gamma^a \nabla^S_\mu \Psi=e^{-\alpha}\gamma^0(\partial_0+\alpha^{\prime})\Psi+ie^{-\alpha} \slashed D\Psi,
\end{align}
where $\slashed D$ denotes the Dirac operator on any of the two-dimensional spatial slices with metric $h_{ij}$ \cite{SGeom}, and the prime stands for the derivative with respect to the conformal time $\eta$.

\subsection{Spectral properties of the Dirac operator in two dimensions and mode expansion}

Let us call ${{\mathcal S}}=\{\Psi\}$ the complex vector space of Dirac spinors, solutions to the Dirac equation derived from the action principle \eqref{action}. This space is equipped with the following positive definite inner product:
\begin{align}\label{inner}
\langle \Psi_1,\Psi_2\rangle=\int \text{d} \tilde\mu \Psi_1^\dagger(\eta,\vec{x}) \Psi_2(\eta,\vec{x}),
\end{align}
where $\text{d} \tilde\mu$ stands for the invariant measure on the spatial manifold provided by the volume element that defines the induced metric $e^{2\alpha}h_{ij}$. This inner product is preserved under evolution on solutions of the Dirac equation \cite{Dimock}, and hence it is independent of the instant of time $\eta$ at which it is evaluated. 

On the other hand, since the Cauchy problem associated with this equation is well posed in globally hyperbolic geometries \cite{Dimock}, as is the case under study, once a spatial section is fixed (by giving a specific value of the time parameter), we can establish an isomorphism between ${\mathcal S}$ and the space of initial data for the Dirac equation. In this sense we can identify both, passing from one to the other in our discussion. In what follows, we will call $S$ this space of initial data.

Besides, in addition to $\Psi$, it is helpful to consider the rescaled field $\psi=e^{\alpha}\Psi$. The reason is that, up to a weight (corresponding to the fourth root of the determinant of the metric $h_{ij}$), the auxiliary field $\psi$ presents symmetric \cite{Casal} canonical Dirac brackets with its corresponding adjoint field \cite{T-N,Dirac} that remain {\sl constant} at all instants of time. We will represent these brackets as an algebra of anticommutators over a Fock space, obtaining in that way a Fock representation of the canonical anticommutation relations and therefore a quantization of both the field $\psi$ and its original unscaled version, $\Psi$.

Thanks to the geodesic completeness of the spatial sections \cite{hopf}, the Dirac operator $\slashed D$ is essentially self-adjoint in the introduced inner product and, in a compact manifold as is our case, it has a discrete spectrum, with eigenvalues $\pm \omega_n$ labeled by natural numbers $n$ such that $\omega_n>0$ grows for increasing $n$ \cite{SGeom}. The space of spinors defined on each of the 2-manifolds then admits a basis formed by the eigenspinors of the Dirac operator. Let us call $\rho^{np}(\vec{x})$ the eigenspinors with positive eigenvalue $\omega_n$, orthonormal in our inner product when this is rescaled by the time dependent factor $e^{-2\alpha}$. The index $p$ accounts for the degeneracy of each eigenspace. We now notice that $\slashed D$ anticommutes with $\gamma^{1}\gamma^{2}=\gamma^{0}$. This is actually a general feature of the Dirac operator on any manifold of even dimension $d$, taking the corresponding generalization of the anticommuting element as $\gamma^{1}...\gamma^{d}$ \cite{SGeom}. Thus we can choose as eigenspinors with negative eigenvalue $-\omega_n$ those defined as $\bar{\sigma}^{np}(\vec{x})=\gamma^0\rho^{np}(\vec{x})$. These eigenspinors automatically form an orthonormal set in the same sense as it happened with $\{\rho^{np}(\vec{x})\}$. 

Let $g_n$ be the degeneracy of each eigenspace, so that, given $n$, $p$ runs from 1 to $g_n$. Its actual value depends on the spectral details of $\slashed D$ and thus on the particular 2-manifold considered. However, for our purposes, we only need its behavior in the ultraviolet regime of large $\omega_n$. Given the counting function $\chi_{\slashed D}(\omega)$ of the Dirac operator on a compact Riemannian manifold of dimension $d$, that counts the number of positive eigenvalues  of $\slashed D$ that are smaller than or equal to $\omega$ (including degeneracy), Weyl's asymptotic formula \cite{JRoe} states that $\chi_\slashed D(\omega)$ grows at most as $\omega^d$ when $\omega$ tends to infinity. In our case, $d=2$, and therefore we conclude that the degeneracy behaves as $g_n=o(\omega_n^2)$ for large $n$, where the symbol $o(\omega_n^2)$ means negligible compared to $\omega_n^2$. 

Having performed the partial gauge fixing explained in the previous section, we can now use the eigenspinors of the Dirac operator on the 2-surfaces to expand the field $\psi$, and therefore any element $\Psi$ of $\mathcal{S}$, as
\begin{align}\label{expan}
\Psi(\eta,\vec{x})=e^{-\alpha(\eta)}\sum_{np}\left[ s_{np}(\eta)\rho^{np}(\vec{x})+\bar{r}_{np}(\eta)\bar{\sigma}^{np}(\vec{x})\right].
\end{align}
We use the bar to denote complex conjugation. In this manner, we have separated the spatial dependence of the spinor from the time dependence. The latter is captured in the {\sl modes}\footnote{Strictly speaking, $s_{np}$ and $\bar{r}_{np}$ are the time-dependent coefficients of the mode expansion of $\psi$.} $s_{np}$ and $\bar{r}_{np}$, which implement the Grassmannian nature of the fermion field. 

Let us translate the field anticommutation relations in terms of the modes $s_{np}$, $\bar{r}_{np}$, and their complex conjugates. Though in the passage to the Hamiltonian formalism, second-class constraints arise for the momenta conjugate to the dynamical modes \cite{Dirac,T-N}, one can easily eliminate them and conclude that the only nonvanishing Dirac brackets are
\begin{align}\label{bra}
\{s_{np},\bar{s}_{np}\}=-i,\qquad  \{r_{np},\bar{r}_{np}\}=-i.
\end{align}
These brackets are symmetric owing to the anticommutativity of our Grassmann variables. Upon quantization, they will become anticommutators of the corresponding operators \cite{Casal}. Notice that, thanks to the rescaling introduced in the fermion field (which extracts a contribution of the scale factor) these brackets do not depend on the background geometry.

\subsection{Fermion dynamics}
\label{dyn}

Introducing the above mode decomposition for the Dirac spinor in the action \eqref{action}, and using the orthonormality of the eigenspinors of the Dirac operator on the spatial slices, we arrive at the expression 
\begin{eqnarray}\label{action2}
I_f&=&\int \text{d}\eta \sum_{np} \Big[ \frac{i}{2} (\bar{s}_{np}s^{\prime}_{np}+{s}_{np}{\bar{s}^{\prime}}_{np}+\bar{r}_{np}r^{\prime}_{np}+{r}_{np}\bar{r}^{\prime}_{np}) \nonumber \\
&+& \omega_n(r_{np}s_{np}+\bar{s}_{np}\bar{r}_{np})-ime^{\alpha}(r_{np}s_{np}-\bar{s}_{np}\bar{r}_{np})\Big].
\end{eqnarray}
By taking Grassmann variational derivatives (e.g. from the left) in the action \eqref{action2}, we get the equations of motion for the fermion modes,
\begin{align}\label{1order}
s_{np}^{\prime}=i(\omega_{n}+ime^{\alpha})\bar{r}_{np}, \qquad r_{np}^{\prime}=-i(\omega_{n}+ime^{\alpha})\bar{s}_{np},
\end{align}
as well as the complex conjugate equations. We observe that the equations of motion are the same for all the modes with the same value of $n$, regardless of the label $p$. These equations can be combined into a second-order differential equation that has the same form for all degenerate modes. Employing $z_{np}$ to denote either $s_{np}$ or $r_{np}$, this second order differential equation reads
\begin{align}\label{2order}
z_{np}^{\prime\prime}=-(\omega_{n}^2+m^{2}e^{2\alpha})z_{np}+i\frac{\alpha^{\prime}me^\alpha}{\omega_n+ime^\alpha}z_{np}^{\prime}.
\end{align}

Solutions to this equation do not depend on the label $p$ and are linear combinations of two complex independent solutions that we will call
\begin{align}\label{sol}
 \Omega^{1}_{n}(\eta)=e^{i\Theta^{1}_{n}(\eta)}, \qquad \Omega^{2}_{n}(\eta)=e^{-i\Theta^{2}_{n}(\eta)}.
\end{align}
Let us set the generic initial conditions $\Theta^{l}_{n}(\eta_{0})=\Theta_{n,0}^{l}$ and $(\Theta^{l}_{n})^{\prime}(\eta_{0})=\Theta^{l}_{n,1}$ for $l=1,2$, at an arbitrary initial time $\eta_{0}$, and let us call 
\begin{align}\label{sol2}
\Omega^{1}_{n,0}=e^{i\Theta^{1}_{n,0}}, \qquad \Omega^{2}_{n,0}=e^{-i\Theta^{2}_{n,0}} . 
\end{align}
Notice that the initial conditions on $\Theta^{l}_{n}$ and their derivatives are related to the initial conditions $s^{0}_{np}$ and $r^{0}_{np}$ on the modes, and on their complex conjugates, via the Dirac equations \eqref{1order}. Taking this into account, one can easily derive the expression for the modes at any time $\eta$, in terms of the two independent solutions of Eq. \eqref{2order} and their initial conditions. Indeed, the time evolution of the modes can be regarded as the following linear transformation of the initial data:
\begin{eqnarray} \label{evol1}
\begin{pmatrix}
s_{np}  \\ \bar{r}_{np} 
\end{pmatrix}_{\!\!\eta}&=&\mathcal{V}_n(\eta,\eta_0)\begin{pmatrix}s_{np}  \\ \bar{r}_{np} 
\end{pmatrix}_{\!\!\eta_0}, \\ \label{evol}
\mathcal{V}_n(\eta,\eta_0)&=&\begin{pmatrix}
\Delta_{n}^{2}e^{i\Theta^{1}_{n}(\eta)}+\Delta^{1}_{n}e^{-i\Theta_{n}^{2}(\eta)} &  \Gamma^{1}_{n}e^{i\Theta_{n}^{1}(\eta)}-\Gamma_{n}^{2}e^{-i\Theta_{n}^{2}(\eta)} \\  \bar{\Gamma}^{2}_{n}e^{i\bar{\Theta}_{n}^{2}(\eta)}-\bar{\Gamma}_{n}^{1}e^{-i\bar{\Theta}_{n}^{1}(\eta)} &
\bar{\Delta}_{n}^{2}e^{-i\bar{\Theta}^{1}_{n}(\eta)}+\bar{\Delta}^{1}_{n}e^{i\bar{\Theta}_{n}^{2}(\eta)}
\end{pmatrix},
\end{eqnarray}
where the subindex $\eta$ in column-vectors means evaluation at that value of the conformal time, and we have defined the constants
\begin{eqnarray}\label{icconsts}
\Delta^{1}_{n}&=&\frac{\Theta_{n,1}^{1}}{\Omega^{2}_{n,0}(\Theta^{1}_{n,1}+\Theta^{2}_{n,1})}, \quad \Delta^{2}_{n}=\frac{\Theta_{n,1}^{2}}{\Omega^{1}_{n,0}(\Theta^{1}_{n,1}+\Theta^{2}_{n,1})},\\ \label{icconsts2} \Gamma_{n}^{l}&=&\frac{\omega_n+im e^{\alpha_0}}{\Omega^{l}_{n,0}(\Theta^{1}_{n,1}+\Theta^{2}_{n,1})}, 
\end{eqnarray}
where $\alpha_{0}=\alpha(\eta_{0})$.

In order to study whether the quantization of the fermion field admits a unitarily implementable dynamics, we do not need the exact solution for the evolution matrix $\mathcal{V}_n(\eta,\eta_0)$, but it suffices to know its behavior in the ultraviolet regime of large $\omega_n$. This analysis is carried out in Appendix \ref{app}. Provided that some regularity conditions apply to the scale factor and its derivatives\footnote{\label{footnote3} Our results apply if the scale factor has a well-defined second derivative with respect to the conformal time that is integrable in every closed interval $[\eta,\eta_0]$.}, in that appendix we conclude that, taking initial conditions $\Omega_{n,0}^{l}=1$ for $l=1,2$, the phases $\Theta^{l}_{n}$ are 
\begin{align}\label{phases}
\Theta^{l}_{n}=\omega_{n}\Delta\eta +\int_{\eta_{0}}^{\eta}\text{d}\tilde\eta\Sigma^{l}_{n}(\tilde\eta),\qquad \Sigma^{l}_{n}(\tilde\eta)=\Lambda^{l}_{n}(\tilde\eta)-\frac{(-1)^l\alpha^{\prime}(\tilde\eta)me^{\alpha(\tilde\eta)} }{2(\omega_n+i m e^{\alpha(\tilde\eta)})},
\end{align}
where $\Delta\eta=\eta-\eta_{0}$, and $\Lambda^{l}_{n}(\tilde\eta)$ is a function that, for large $\omega_n$, is at most of order $\omega_n^{-1}$: $\Lambda^{l}_{n}\sim\mathcal{O}(\omega_n^{-1})$.
On the other hand, we also obtain
\begin{align}\label{constants}
\Delta^{l}_{n}=\frac12-(-1)^l \frac{\alpha^{\prime}_0 m e^{\alpha_0}}{4\omega_n(\omega_n+i m e^{\alpha_0})},\qquad \Gamma^{l}_{n}=\frac12+i\frac{me^{\alpha_0}}{2\omega_n}.
\end{align} 

\section{Fermion complex structures}
\label{sec:CS}

\subsection{Ambiguity in the Fock quantization} 

We can characterize the Fock representation of the canonical anticommutation relations by means of the mathematical object known as complex structure. In its covariant version, a complex structure ${\mathcal J}$ is defined as a real linear map on the space of solutions $\mathcal{S}$, the square of which is minus the identity, and such that it is compatible with the inner product, namely $\langle \mathcal{J}\Psi_1,\mathcal{J}\Psi_2\rangle=\langle\Psi_1,\Psi_2\rangle$ \cite{wald,Smatrix}. Using the existing isomorphism between ${\mathcal S}$  and the space of initial data $S$ once a reference time $\eta_0$ is given, we can equivalently introduce a canonical complex structure $J$ on $S$, corresponding to $\mathcal J$. Thus, in our following discussion, any consideration in the covariant approach can be straightforwardly translated to the canonical language, and vice versa, as long as we fix the (otherwise arbitrary) Cauchy reference surface for the identification \cite{unit}.

A complex structure $\mathcal J$ defines a splitting of ${\mathcal S}$ into two mutually complementary and orthogonal subspaces, ${\mathcal S}^\pm_{\mathcal J}=({\mathcal S}\mp i {\mathcal J}{\mathcal S})/2$, where ${\mathcal S}_{\mathcal J}^\pm$ are the eigenspaces of ${\mathcal J}$ with eigenvalues $\pm i$, respectively. Analogously, we can define the complex structure ${\mathcal J}$ as a linear map on the complex conjugate space $\bar{\mathcal S}$, for which the inner product is the complex conjugate of Eq. \eqref{inner}. The complex structure then induces a splitting equivalent to the previous one, with $(\bar{{\mathcal S}}_{\mathcal J})^\pm=(\bar{{\mathcal S}}\mp i {\mathcal J}\bar{{\mathcal S}})/2=\overline{({\mathcal S}_{\mathcal J}^\mp)}$. These splittings are in turn associated with a specific choice of creation and annihilationlike variables (to be represented as creation and annihilation operators), and therefore with a specific definition of particles and antiparticles. Indeed, the subspace ${\mathcal S}^{+}_{\mathcal J}$ provides the particle annihilationlike part of the Dirac spinors, whereas ${\mathcal S}^{-}_{\mathcal J}$ corresponds to the antiparticle creationlike part. The Fock representation associated with ${\mathcal J}$ is completely characterized by the assignation of the one-particle Hilbert space of particles to the completion of ${\mathcal S}_{\mathcal J}^+$ with the inner product \eqref{inner}, and the assignation of the one-particle Hilbert space of antiparticles to the completion of $\overline{({\mathcal S}^-_{\mathcal J})}$. The direct sum of these two Hilbert spaces gives the one-particle Hilbert space of the representation, from which the antisymmetric Fock space is constructed \cite{wald}. This representation is alternatively specified by its vacuum state, defined as the normalized cyclic state which vanishes under the action of all the annihilation operators. It is clear that different choices of complex structures provide different one-particle Hilbert spaces, with different vacua, which may in general be inequivalent to one another, thus defining physically distinct quantum theories. We will show how our criterion is capable to remove this ambiguity in the Fock representation of the canonical anticommutation relations of the Dirac field.

A related ambiguity lies in the freedom of choosing a specific set of creation and annihilationlike variables to represent the field among all those that are related by back\-ground-dependent linear transformations that affect the time dependence. Indeed, one may always absorb part of the field evolution into an explicit dependence of the field on the scale factor, the function of the background geometry that encodes its variation in time. In this way, one can change the dynamics of the family of creation and annihilationlike variables that are going to be promoted to operators, and therefore modify the feasibility of the requirement of unitary implementability of their evolution, a key ingredient in our quantization criterion. 

In the following we will focus on {\emph{invariant} complex structures, that is to say,}  complex structures that commute with the action of the symmetry group of the equations of motion for the modes, so that they lead to Fock representations with vacua that are invariant under the unitary transformations generated by those symmetries.

\subsection{Invariant vacua}
\label{ann-cre}

In Sec. \ref{dyn} we saw that the equations of motion for the fermion modes only couple the modes $s_{np}$ and $\bar{r}_{np}$, for the same labels $n$ and $p$, and similarly for their complex conjugates. Moreover, those equations do not depend on the degeneracy label $p$. In consequence, the field equations are invariant under all transformations that interchange eigenmodes of the Dirac operator without affecting the value of $\omega_n$. In general, these transformations need not be in correspondence with Killing symmetries of the spatial sections (with metric $h_{ij}$), although it is clear that any such spatial isometry of the system is in particular a symmetry of the equations of motion, because the Dirac operator is built from the spatial metric $h_{ij}$. For cases with maximally symmetric spatial manifolds, there are enough Killing symmetries available already at the level of the spatial geometry, and one can consider just the corresponding group of isometries to restrict the quantization.

After these clarifications, it should be obvious that any linear transformation that commutes with the action of the group of symmetries of the Dirac equation must diagonalize into $2\times 2$ blocks: these blocks can at most mix the modes $s_{np}$ and  $\bar{r}_{np}$ with the same value of $p$. Besides, if we appeal to all the symmetry at hand, the blocks must be equal for all the modes associated with the same eigenvalue of the Dirac operator (in norm). In this sense, invariant complex structures are totally determined by a series of $2 \times 2$ matrices labeled by the integer $n\in \mathbb{N}$.

Since an invariant complex structure can only mix the modes $s_{np}$ and  $\bar{r}_{np}$ with the same labels, the associated creation and annihilationlike variables, that diagonalize its action, will be given by linear combinations of these modes. We will denote by $a_{np}$ and $b_{np}$ the annihilationlike variables of the particles and antiparticles, respectively, while the creationlike variables are their complex conjugates, that is, $a_{np}^{\dagger}=\bar{a}_{np}$ and $b_{np}^{\dagger}=\bar{b}_{np}$. These variables need to satisfy the characteristic Dirac brackets for creation and annihilation sets, 
\begin{align}\label{CACR}
\{a_{np},a^{\dagger}_{np}\}=
\{b_{np},b^{\dagger}_{np}\}=-i,\qquad \{a_{np},b_{np}\}=0.
\end{align}

Let us now consider time-dependent families of creation and annihilationlike variables associated with invariant complex structures. For each value of the conformal time, the creation and annihilationlike variables are linear combinations of the modes $s_{np}$ and  $\bar{r}_{np}$, as we have seen, but now the combination will be allowed to vary from an instant of time to another. More specifically, we will have a linear relation of the form
\begin{align}\label{blockcs1}
\begin{pmatrix}
a_{np}  \\ b^{\dagger}_{np} 
\end{pmatrix}_{\!\!\eta}=\mathcal{F}_n(\eta)
\begin{pmatrix}
s_{np}  \\ \bar{r}_{np} 
\end{pmatrix}_{\!\!\eta},\qquad \mathcal{F}_n(\eta)=\begin{pmatrix}
f_1^n(\eta) & f_2^n(\eta) \\ g_1^n(\eta) & g_2^n(\eta)
\end{pmatrix}.
\end{align}
The time-dependent functions $f_{l}^{n}$ and $g_{l}^{n}$ ($l=1,2$), together with  their complex conjugates $\bar{f}^{n}_{l}$ and $\bar{g}^{n}_{l}$, are taken to be continuous (and as smooth as required for a suitable evolution of the classical  creation and annihilationlike variables) and must satisfy the relations
\begin{align}\label{sympl}
|f_{1}^{n}|^{2}+|f_{2}^{n}|^{2}=1, \qquad |g_{1}^{n}|^{2}+|g_{2}^{n}|^{2}=1, \qquad f_{1}^{n}\bar{g}^{n}_{1}+f_{2}^{n}\bar{g}^{n}_{2}=0,
\end{align}
so that Eq. \eqref{CACR} is verified. 

If we interpret the resulting families of creation and annihilationlike variables as a dynamical trajectory, the time dependence introduced in the linear combinations changes the evolution from that of the modes $s_{np}$ and  $\bar{r}_{np}$ (which is just the evolution of the field $\psi$); indeed, part of that evolution is now compensated by the explicit dependence on $\eta$ of the matrices $\mathcal{F}_n$.

Conditions \eqref{sympl} guarantee the invertibility of the relation between the creation and annihilationlike variables (on the one hand) and the modes $s_{np}$ and  $\bar{r}_{np}$ (on the other hand) at all times. Substituting this inversion in the expansion \eqref{expan}, one obtains the expression of the Dirac field in terms of the introduced family of variables. From that expression, it is then obvious that the dynamics of the field $\Psi$ is not given exclusively by the evolution of the creation and annihilationlike variables, but in addition the field has explicit time-dependent contributions coming from $\mathcal{F}_n$ and from the rescaling by the scale factor that relates $\Psi$ and $\psi$. It is only the part corresponding to the evolution of the creation and annihilationlike variables the one that we want to implement quantum mechanically.

The challenge that we have to face in the rest of the work is to show that, once the invariance under the symmetries of the equations of motion has been imposed, the requirement of a  unitarily implementable dynamics for our family of creation and annihilationlike variables indeed is enough to select a nontrivial quantum evolution for the fermion field as well as to pick out a unique preferred Fock representation for the system (up to unitary transformations). 

As we will discuss in detail in the next section, the linearity of the evolution of the creation and annihilationlike variables allows us to express them at any time in terms of their initial values at a spatial surface, that we fix from now on at constant conformal time $\eta_0$. For those initial values at $\eta_0$, we can take precisely the values $\{a_{np}(\eta),b_{np}(\eta)\}$ \emph{at time} $\eta$, as well as their complex conjugates, in a process that corresponds to letting these variables evolve from $\eta_0$ to $\eta$ and identify their values at $\eta$ with new initial conditions at $\eta_0$. We can then consider the complex structures that have a diagonal action precisely on these sets of variables at the reference initial surface, either adopting a straightforward canonical picture, or in the equivalent covariant language (regarding the initial values as constants in an expansion in a basis of solutions). In this way, we will arrive to a family of complex structures ${\mathcal J}_{\eta}^{(\eta_0)}$ (or $J_{\eta}^{(\eta_0)} $), parametrized by the conformal time and defined for an arbitrary initial time $\eta_0$, that can be regarded as the family induced by the evolution of the creation and annihilationlike variables selected for the description of the field. This family of complex structures will be all equivalent, and therefore will not involve any physical ambiguity for the Fock representation of the field, if and only if the evolution of the creation and annihilationlike variables that we have chosen is implementable quantum mechanically as a unitary transformation. 

Finally, before discussing this implementability, it is worth pointing out that, combining conditions \eqref{sympl}, one can write
\begin{align}\label{fgrel}
g_{1}^{n}=\bar{f}_{2}^{n}e^{iG^{n}}, \qquad g_{2}^{n}=-\bar{f}^{n}_{1}e^{iG^{n}},\qquad f_{1}^{n}g^{n}_{2}-g^{n}_{1}f^{n}_{2}=-e^{iG^{n}},
\end{align}
with $G^{n}$ a certain phase. Hence, only one of the four functions $\{f_{l}^{n},g_{l}^{n}\}$, together with two additional phases for each $n$, suffice to characterize the considered family of creation and annihilationlike variables.

\section{Unitarity of the evolution}
\label{sec:Unitarity}

Let us restrict our attention to quantizations that satisfy the uniqueness criteria employed in the case of four-dimensional spacetimes \cite{compact,compact2,unit,uf}, namely those that are not only \emph{invariant}, but also allow for a unitary implementability of our nontrivial dynamics in the quantum theory. 

For any of the allowed families of creation and annihilationlike variables, introduced in the previous section, the dynamical evolution is a Bogoliubov transformation, formed by a collection of matrices $\mathcal{B}_{n}(\eta,\eta_{0})$, that relates the considered variables at two different times (say, the arbitrary time $\eta$ and the initial time $\eta_0$):
\begin{align} \label{bog}
\begin{pmatrix} a_{np} \\ b^{\dagger}_{np} \end{pmatrix}_{\!\!\eta}=\mathcal{B}_{n}(\eta,\eta_{0})\begin{pmatrix} a_{np} \\ b^{\dagger}_{np} \end{pmatrix}_{\!\!\eta_{0}}, \qquad \mathcal{B}_{n}(\eta,\eta_{0})=\begin{pmatrix} \alpha_{n}^{f}(\eta,\eta_{0}) & \beta_{n}^{f}(\eta,\eta_{0}) \\ \beta_{n}^{g}(\eta,\eta_{0}) & \alpha_{n}^{g}(\eta,\eta_{0}) \end{pmatrix}.
\end{align}
The transformation \eqref{bog} is implementable as a unitary operator in the Fock space defined by the complex structure ${\mathcal J}_{\eta_{0}}^{(\eta_0)}$ (introduced at the end of Sec. \ref{ann-cre}) if and only if its antilinear part is Hilbert-Schmidt \cite{Shale}, that is to say, if and only if
\begin{align}\label{ucond}
\sum_{n}g_{n}|\beta_{n}^{f}(\eta,\eta_0)|^{2}<\infty \qquad \text{and} \qquad \sum_{n}g_{n}|\beta_{n}^{g}(\eta,\eta_0)|^{2}<\infty,
\end{align}
for all times $\eta$ in the evolution, where we have taken into account the degeneracy $g_n$ of the Dirac eigenspaces.

Recall that, in our case of compact spatial two-dimensional sections, Weyl's asymptotic formula \cite{JRoe} tells us that the number  $\chi_{\slashed D}(\omega)$ of eigenstates of the Dirac operator with positive eigenvalue smaller than or equal to $\omega$ grows asymptotically as $\omega^2$. Therefore, the number of eigenstates with eigenvalue between $N$ and $N-1$ grows, for a large integer $N$, as $N^2-(N-1)^2$, i.e., asymptotically as $2N$. Thus, for sufficiently large $N$, there exists a positive constant $K$ such that this number of eigenstates is bounded by $K N$. Employing this behavior and the monotonicity of the decreasing function $\omega^{-2}$ for positive $\omega$, we see that, for large $N_0$ and arbitrarily large $N_1>N_0$, and with $N_0\geq \omega_{n_0}>N_0-1$ and $N_1\geq \omega_{n_1}>N_1-1$,
\begin{align}\label{omegatwo} \sum_{n=n_0}^{n_1} \frac{g_n} {\omega_n^4} \leq \sum_{N=N_0}^{N_1} \frac{\chi_{\slashed D}(N)- \chi_{\slashed D}(N-1)}{(N-1)^4}\leq \sum_{N=N_0}^{N_1} \frac{K}{N^3}.
\end{align}
Since the sum on the right-hand side converges when $N_1$ goes to infinity, we can assure that the sequence $\{\sqrt{g_n}\omega_n^{-2}\}_{n\in {\mathbb{N}}}$ is square summable, and hence also any subsequence of it. 

Using Eqs. \eqref{evol1}, \eqref{blockcs1}, and \eqref{bog}, we obtain the explicit form of the Bogoliubov transformation: $\mathcal{B}_{n}(\eta,\eta_{0})=\mathcal{F}_n(\eta) \mathcal{V}_n(\eta,\eta_0) \mathcal{F}^{-1}_n(\eta_0)$. Then, the quantum analog of Eq. \eqref{bog}, attained by promoting the creation and annihilationlike variables to operators, is by construction a solution to the Heisenberg's equations associated precisely with those creation and annihilation operators. The unitary implementability of this quantum transformation therefore amounts to the possibility of finding a unitary operator that, when acting on those creation and annihilation operators at the initial time $\eta_0$, reproduces exactly the linear evolution determined by Eq. \eqref{bog}. In order to discuss this unitarity, we use the form of the beta coefficients of the considered Bogoliubov transformation, which are given by
\begin{align}
\beta_{n}^{h}=&-\frac{1}{h_{1}^{n,0}k_{2}^{n,0}-h_{2}^{n,0}k_{1}^{n,0}}
\bigg[\big(\Delta_{n}^{2}h_{2}^{n,0}-\Gamma^{1}_{n}h_{1}^{n,0}\big)h_{1}^{n}e^{i\Theta_{n}^{1}}+\big(\Delta_{n}^{1}h_{2}^{n,0}+\Gamma^{2}_{n}h_{1}^{n,0}\big)h_{1}^{n}e^{-i\Theta_{n}^{2}} \nonumber \\ 
-&\big(\bar{\Delta}_{n}^{1}h_{1}^{n,0}-\bar{\Gamma}^{2}_{n}h_{2}^{n,0}\big)h_{2}^{n}e^{i\bar{\Theta}_{n}^{2}}-\big(\bar{\Delta}_{n}^{2}h_{1}^{n,0}+\bar{\Gamma}^{1}_{n}h_{2}^{n,0}\big)h_{2}^{n}e^{-i\bar{\Theta}_{n}^{1}}\bigg].\label{beta}
\end{align}
Here, $\{h,k\}=\{f,g\}$ as a set, with $h$ being equal to either $f$ or $g$ and $k$ being the complementary of $h$. We have omitted the dependence of these functions on $\eta$ to alleviate the notation and denoted the evaluation at $\eta_{0}$ with the superscript $0$ (preceded by a comma). 

Now, taking into account Eqs. \eqref{phases} and \eqref{constants}, as well as relations \eqref{sympl} and \eqref{fgrel}, we find the following expression for the norm of the beta coefficients in the ultraviolet regime:
\begin{align}
|\beta_{n}^{h}|=\frac12
\bigg|&\big(h_{2}^{n,0}-h_{1}^{n,0}\big)\left[h_{1}^{n}\left(1+i\int \Sigma^1_n\right)+h_{2}^{n}\left(1+i\int \bar\Sigma^2_n\right)\right] e^{i\omega_{n}\Delta\eta}\nonumber \\+&\big(h_{2}^{n,0}+h_{1}^{n,0}\big)\left[h_{1}^{n}\left(1-i\int \Sigma^2_n\right)-h_{2}^{n}\left(1-i\int \bar\Sigma^1_n\right)\right]e^{-i\omega_{n}\Delta\eta}\nonumber\\
+&\frac{2me^{\alpha_0}}{\omega_n}\left(h_{1}^{n,0}h_{1}^{n}+h_{2}^{n,0}h_{2}^{n}\right)\sin(\omega_n\Delta\eta)\bigg|+\mathcal{O}(\omega_n^{-2}),\label{beta3}
\end{align}
where the integrals are in conformal time, over the interval $[\eta_{0},\eta]$. 

Conditions \eqref{ucond} impose a very concrete behavior of the coefficients $h_{l}^{n}$, both in their time and mode dependence. Let us emphasize that we are interested in the unitary implementability of the Bogoliubov transformation \eqref{bog}, that provides the evolution of the creation and annihilation operators, but without trivializing it. Obviously, one can always adapt the explicit time dependence introduced in the definition \eqref{blockcs1} of our creation and annihilationlike variables so that it absorbs the dynamical variation experienced by the field modes $\{s_{np},\bar{r}_{np}\}$, making in this way $\mathcal{B}_n(\eta,\eta_0)$ the identity matrix. Nevertheless, in our globally hyperbolic manifold, the criterion of a unitary implementation of the identity transformation is useless, since it poses no restriction on the Fock representation. Moreover, we may adopt similar choices for $\mathcal{F}_n(\eta)$ that, by rendering $\mathcal{B}_n(\eta,\eta_0)$ equal to the identity only up to terms that are sufficiently negligible in the asymptotic limit of large $\omega_n$, would not compromise the trivial unitary implementability. By inspection of Eq. \eqref{beta3}, one can convince oneself that the dominant contribution of the dynamics of the modes $\{s_{np},\bar{r}_{np}\}$ is in fact given by (imaginary  exponentials of) the phases $\pm \omega_n \Delta \eta$. Hence, in order to avoid a trivial evolution, we rule out the possibility that these dynamical contributions are counterbalanced with a specific choice of time and mode-dependent phases in the linear combinations that determine our creation and annihilationlike variables. Indeed, this kind of strategy allows one to single out a unitarily implementable quantum evolution (modulo unitary redefinitions) in the case of scalar fields and fermion fields in 3+1 dimensions, for different backgrounds with maximally symmetric, compact spatial sections (see e.g. \cite{compact2,uf}).

Let us then analyze the restrictions imposed by the unitarity conditions \eqref{ucond}, studying first the case in which the fermion field is massive. Specifically, taking into account relations \eqref{sympl} and \eqref{fgrel}, the following scenarios can be distinguished (at least in a sufficiently short interval beyond $\eta_{0}$):
\begin{itemize}
\item[i)] $h_{l}^{n}$ tends to zero in the asymptotic limit of large $\omega_{n}$ for an infinite subset of the natural numbers, $n\in\mathbb{N}^{\downarrow}_{l}$. 

Let us introduce the notation $\{l,\tilde{l}\}= \{1,2\}$, as a set. Since $|h_{\tilde{l}}^{n}|^{2}+|h_{l}^{n}|^{2}=1$, we can write
\begin{align}\label{hs}
h_{\tilde{l}}^{n}=e^{iH^{n}_{\tilde{l}}}\sqrt{1-|h_{l}^{n}|^{2}},
\end{align}
with $H^{n}_{\tilde{l}}$ being some possibly time-dependent phase. Thus, we see that $h_{\tilde{l}}^{n}$ is of order one. Besides, at leading order we obtain that
\begin{align}
|\beta_{n}^{h}|= \left|h_{\tilde l}^{n,0}h_{\tilde l}^{n} \sin(\omega_{n}\Delta\eta)\right| +o(1),\quad\forall n\in\mathbb{N}^{\downarrow}_{l},
\end{align}
where we recall that the symbol $o(\,\,)$ means negligible with respect to its argument. Therefore, in this scenario, the norm of an infinite number of beta coefficients is of order one, so that conditions \eqref{ucond} fail to hold; i.e. time evolution is not unitarily implementable.

\item[ii)] Neither $h_{1}^{n}$ nor $h_{2}^{n}$ tend to zero in the asymptotic limit of large $\omega_{n}$, for all $n\in\mathbb{N}$ greater than a certain $\tilde{n}\geq 0$.

A necessary condition to attain the square summability of the sequences $\{\sqrt{g_n}\beta_{n}^{h}\}_{n\in\mathbb{N}}$ is that there is no contribution of unit order to the norm of any infinite subset of beta coefficients. At this stage, there are two possibilities (that are not mutually exclusive): either $h_2^n=\pm h_1^n$ up to terms negligible compared to the unit, or 
\begin{align}
 \frac{h_2^n+h_1^n}{h_2^n-h_1^n}=\frac{h_{2}^{n,0}+h_{1}^{n,0}}{h_{2}^{n,0}-h_{1}^{n,0}}e^{-2i\omega_{n}\Delta\eta}+o(1),
\end{align}
for all $\eta$ and for all $n>\tilde{n}$. The latter possibility can be regarded as a trivialization of the dynamics, achieved by absorbing the dominant dynamical variation of the phases of the fermion modes in the choice of canonical variables for the description of the field. This requires a very specific fixation of the phases of $h_{1}^{n}$ and $h_{2}^{n}$ that we will rule out, e.g. by demanding that the dominant contribution in our coefficients $h_{l}^{n}$ be mode independent. More generally, in all of our following considerations and statements of results, and according to our previous comments, we will automatically discard a time dependence of this sort that would trivialize the dynamics of the Dirac field. Therefore, we only consider as allowed the former possibility, so that asymptotically
\begin{eqnarray}\label{hh}
h_{l}^{n}&=&\frac{e^{iH^{n}_{l}}}{\sqrt{2}}+\vartheta^{n}_{h,l}, \\ \label{hh2} h_{\tilde{l}}^{n}&=&\pm e^{iH^{n}_{l}}\sqrt{1-|h_{l}^{n}|^{2}}=\pm e^{iH^{n}_{l}} \left[\frac{1}{\sqrt{2}}-\Re(e^{-iH^{n}_{l}}\vartheta^{n}_{h,l})\right]+\mathcal{O}(|\vartheta^{n}_{h,l}|^{2}),
\end{eqnarray}
for an infinite subset of the natural numbers, $n\in\mathbb{N}^{\pm}_{l}$, where $\vartheta^{n}_{h,l}$ is some complex, time-dependent and mode-dependent function that tends to zero in the large $\omega_{n}$ limit. Here, we are thinking on the natural numbers as given by four subsets, $\mathbb{N}=\mathbb{N}^{+}_1\cup\mathbb{N}^{-}_1\cup\mathbb{N}^{+}_2\cup\mathbb{N}^{-}_2$ (up to a finite number of elements), allowing for the possibility that up to three of these subsets be void, and considering that $h_{l}^{n}$ be  $h_{1}^{n}$ for $n\in\mathbb{N}^{\pm}_{1}$ and $h_{2}^{n}$ for $n\in\mathbb{N}^{\pm}_{2}$, with the $\pm$ superscripts indicating the relative sign for $h_{\tilde{l}}^{n}$ in Eq. \eqref{hh2}. Besides, we have taken into account conditions \eqref{sympl} in order to fix the leading order in both coefficients, as well as the subleading terms in $h_{\tilde{l}}^{n}$. Then, for all $n\in\mathbb{N}^{\pm}_{l}$, the norm of the corresponding beta coefficients is asymptotically given by
\begin{align}\label{betass}
|\beta_{n}^{h}|=
\frac{1}{\sqrt{2}}\bigg|& \left[\vartheta^{n,0}_{h,l}e^{-iH^{n,0}_{l}}+\Re(e^{-iH^{n,0}_{l}}\vartheta^{n,0}_{h,l})-i(-1)^{l}\frac{me^{\alpha_{0}}}{\sqrt{2}\omega_n}\right]e^{ \pm i \omega_{n}\Delta\eta }\nonumber\\
&
- \left[\vartheta^{n}_{h,l}e^{-iH^{n}_{l}}+\Re(e^{-iH^{n}_{l}}\vartheta^{n}_{h,l})-i (-1)^{l}\frac{me^{\alpha}}{\sqrt{2}\omega_n}\right]e^{\mp  i \omega_{n}\Delta\eta}
\bigg|
\end{align}
up to terms that are either negligible compared to $\vartheta^{n}_{h,l}$ or to $\omega_{n}^{-1}$, whichever is larger. To obtain this expression we have used that
\begin{align}\label{sig}
\int \left(\bar\Sigma^1_n-\Sigma^2_n\right)=\frac{m}{\omega_n}\left(e^\alpha-e^{\alpha_0}\right)+\mathcal{O}(\omega_n^{-2}),
\end{align}
a result that is proven in Appendix \ref{appb}. Two different situations might occur at this point, depending on the spectral properties of the Dirac operator on the spatial slices, and on the nature of the considered subsets of $\mathbb{N}$: either the sequence $\{\sqrt{g_{n}}\omega_{n}^{-1}\}_{n\in\mathbb{N}^{\pm}_{l}}$ is square summable, or it is not. If the first situation happened to be the case, it is not hard to convince oneself that the dynamical beta coefficients would be square summable (including degeneracy) in the subset $\mathbb{N}_{l}^{\pm}$, for all times $\eta$, if and only if the sequence $\{\sqrt{g_{n}}\vartheta^{n}_{h,l}\}_{n\in\mathbb{N}^{\pm}_{l}}$ is square summable as well.

Let us now study the alternative situation, that happens if $\{\sqrt{g_{n}}\omega_{n}^{-1}\}_{n\in\mathbb{N}^{\pm}_{l}}$ fails to be a square summable sequence. In this case, the summability conditions \eqref{ucond}, restricted to any of the subsets $\mathbb{N}^{\pm}_{l}$, where they must in particular be satisfied, imply that the function $\vartheta^{n}_{h,l}$ has to be such that
\begin{align}\label{theta}
\vartheta^{n}_{h,l}+e^{iH^{n}_{l}} \Re(e^{-i H^{n}_{l}}\vartheta^{n}_{h,l})-i (-1)^{l}\frac{me^{\alpha}}{\sqrt{2}\omega_n}e^{iH^{n}_{l}}=o(\omega_{n}^{-1}).
\end{align}
Let us call $\tilde \vartheta^{n}_{h,l}$ the expression in the left-hand side of the above equation, which must be negligible compared to $\omega_{n}^{-1}$. Then, by introducing the resulting behavior of the functions $h^n_l$ and $h_{\tilde{l}}^{n}$ in the norm of the beta coefficients, we get 
\begin{align}
|\beta_{n}^{h}|=
\frac{1}{\sqrt{2}}\bigg|\tilde \vartheta^{n,0}_{h,l}e^{-iH^{n,0}_{l}} e^{\pm  i\omega_{n}\Delta\eta}
-\tilde \vartheta^{n}_{h,l}e^{-iH^{n}_{l}} e^{\mp i\omega_{n}\Delta\eta}\bigg|+\mathcal{O}(\omega_{n}^{-2}).
\end{align}
Therefore, given the functional independence of the exponentials in time, together with the implications of Eq. \eqref{omegatwo} coming from the asymptotic behavior of the degeneracy $g_{n}$ (and recalling that we exclude a trivialization of the fermion dynamics), we conclude in this situation that conditions \eqref{ucond} are verified in any of the subsets $\mathbb{N}^{\pm}_{l}$ if and only if the sequence $\{\sqrt{g_{n}}\tilde \vartheta^{n}_{h,l}\}_{n\in\mathbb{N}^{\pm}_{l}}$ is square summable for all times $\eta$.

\end{itemize}

It is not hard to see, using relations \eqref{sympl}, that, once $h_{l}^{n}$ is fixed as either $f_{l}^{n}$ or $g_{l}^{n}$, then the other beta functions, which can be called $\beta^{k}_{n}(\eta,\eta_{0})$ with our notation, have the same norm as $\beta^{h}_{n}(\eta,\eta_{0})$. So, if the first condition in Eq. \eqref{ucond} is true, then the second one also holds, and the dynamics of the selected creation and annihilationlike variables is unitarily implementable in the Fock space.

In summary, the necessary and sufficient conditions for an invariant Fock representation of the canonical anticommutation relations of the massive Dirac field to support the definition of a unitarily implementable nontrivial quantum dynamics are that the corresponding functions $h_{l}^{n}$ and $h_{\tilde l}^{n}$ be asymptotically of the form \eqref{hh} and \eqref{hh2} for $n\in \mathbb{N}^{\pm}_l$, and the terms $\vartheta^{n}_{h,l}$ either: a) form a sequence that is square summable including the degeneracy, if the sequence $\{\sqrt{g_{n}}\omega_{n}^{-1}\}_{n\in\mathbb{N}^{\pm}_{l}}$ happened to be itself square summable, or otherwise b) they satisfy Eq. \eqref{theta}, with square summable subleading terms. We recall that we have split the natural numbers, up to a finite collection of them, into four infinite subsets $\mathbb{N}^{\pm}_{1}$ and $\mathbb{N}^{\pm}_{2}$, allowing the possibility that up to three of these subsets be void. Our statement is true as long as the different time-dependent phases in the definition of the creation and annihilationlike variables are not fixed so as to trivialize the dominant dynamical contribution in the phases of the fermion modes, with respect to an asymptotic expansion in terms of $\omega_{n}$. 

To conclude this section, let us briefly discuss the unitary implementability of the dynamics in the particular case in which the mass of the fermion field is zero. This situation can be regarded as special, inasmuch as it displays conformal invariance and admits a ``unique'' natural vacuum that does not evolve with respect to the mode variables $s_{np}$ and $\bar{r}_{np}$: the conformal vacuum\footnote{Notice that, nonetheless, the evolution of the Dirac field still contains an explicitly time-dependent part, given by the rescaling introduced to pass from $\Psi$ to $\psi$, as well as the quantum dynamical part corresponding to the mode variables $s_{np}$ and $\bar{r}_{np}$.}. In fact, in this particular case, the norm of the beta coefficients simplifies to
\begin{align}
|\beta_{n}^{h}|=\frac{1}{2} \bigg|&\big(h_{2}^{n,0}-h_{1}^{n,0}\big)\big(h_{1}^{n}+h_{2}^{n}\big) e^{i\omega_{n}\Delta\eta}+\big(h_{2}^{n,0}+h_{1}^{n,0}\big)\big(h_{1}^{n}-h_{2}^{n}\big)e^{-i\omega_{n}\Delta\eta}\bigg|,
\end{align}
so that, following the same line of reasoning as in the massive case, the families of creation and annihilationlike variables selected by our criteria of symmetry invariance and unitarity can be completely characterized by coefficients $(h^{n}_{l},h^{n}_{\tilde l})$ of the asymptotic form \eqref{hh} and \eqref{hh2} for all $n\in \mathbb{N}^{\pm}_l$, and such that the sequences  $\{\sqrt{g_n} \vartheta^{n}_{h,l}\}_{n\in\mathbb{N}^{\pm}_l}$ are square summable. Within this family, the simple choice $f_1^n=f_2^n=g_{1}^n=-g_2^n=1/\sqrt{2}$ defines a representation for which the beta coefficients identically vanish. This representation corresponds precisely to the mentioned conformal vacuum.

Remarkably, in both the massless and the massive cases, the criterion of unitary implementability of the dynamics fixes completely (up to phases) the leading order (which is of order unity) as well as the term of order $\omega_n^{-1}$, if it is present, in the asymptotic expansion of the coefficients $(f^{n}_{l},f^{n}_{\tilde l})$ and $(g^{n}_{l},g^{n}_{\tilde l})$. Indeed, the leading order in that expansion may depend on time and on the mode only through a phase, and has constant norm equal to $1/\sqrt{2}$. A term of order $\omega_n^{-1}$ does not show up in the massless case, whereas for the massive case this next-to-leading order term may be needed for a unitarily implementable dynamics. If this happens, its contribution vanishes either for the coefficient $h_1^n$ or for $h_2^n$. Besides, this additional term, if necessary, must equal the function $e^\alpha m/(\sqrt{2}\omega_n)$ in norm. The coefficients for the massive case thus depend in a very precise way, not only on the eigenvalue of the Dirac operator, but also on the mass of the field and, more importantly, on time, via a dependence on the background where the field propagates. 

\section{Uniqueness of the Fock representation}
\label{proof}

In the previous section we characterized all families of creation and annihilationlike variables for the Dirac field that, sharing the symmetries of the equations of motion, evolve according to nontrivial dynamics that are unitarily implementable quantum mechanically. Each of these families determine both a family of unitarily equivalent complex structures and a notion of quantum dynamics that relates them. This is the same as determining a Fock representation (e.g. that associated with the choice of creation and annihilationlike variables at the initial time $\eta_0$) and a quantum evolution in the corresponding Fock space. In the following, we will refer to this combination of a Fock representation and a specific quantum dynamics as a {\sl Fock quantization} of the system. 

In this section, we prove that, once the convention for particles and antiparticles has been settled, all invariant Fock quantizations with unitarily implementable dynamics are, in fact, unitarily equivalent. In order to prove this result, let us introduce as \emph{reference} Fock quantization the one characterized by a choice of creation and annihilationlike variables with
\begin{align}\label{reference}
f_{1}^{n}=\frac{1}{\sqrt{2}}\left(1-i\frac{e^\alpha m}{\omega_n}\right),\qquad f_{2}^{n}=\sqrt{1-|f_{1}^{n}|^{2}},\qquad g_{1}^{n}=f_{2}^{n}, \qquad g_{2}^{n}=-\bar{f}_{1}^{n}.
\end{align}
The above choice is one of the simplest that satisfy our requirements. In the case of the massless field, it defines the {\it natural} quantization with conformal vacuum. Now, let us call $\{\tilde{a}_{np}^{\dagger},\tilde{b}_{np}^{\dagger},\tilde{a}_{np},\tilde{b}_{np}\}$ any other choice of creation and annihilationlike variables that defines a Fock quantization with a nontrivial and unitarily implementable dynamics. These variables will be characterized by coefficients $\tilde{f}_{l}^{n}$ and $\tilde{g}_{l}^{n}$, as those given in Eqs. \eqref{sympl} and \eqref{fgrel}, that are of the asymptotic form \eqref{hh}, and such that the corresponding sequences formed by $\vartheta^{n}_{\tilde h,l}$ either satisfy Eq. \eqref{theta} with a square summable left-hand side, or are square summable including degeneracy, in case that $g_{n}\omega_{n}^{-2}$ were summable over $\mathbb{N}^{\pm}_{l}$.

Given the reference Fock quantization and another arbitrary one allowed by our criterion, the creation and annihilationlike variables associated with them are related via a time-dependent Bogoliubov transformation,
\begin{align}\label{bogK}
\begin{pmatrix} \tilde a_{np} \\ \tilde b^{\dagger}_{np} \end{pmatrix}_{\!\!\eta}=\mathcal{K}_{n}(\eta)\begin{pmatrix} a_{np} \\ b^{\dagger}_{np} \end{pmatrix}_{\!\!\eta},\qquad  \mathcal{K}_{n}=\begin{pmatrix} \kappa_{n}^{f} & \lambda_{n}^{f} \\ \lambda_{n}^{g} & \kappa_{n}^{g} \end{pmatrix}.
\end{align}
As before, this Bogoliubov transformation, formed by the sequence of matrices $\mathcal{K}_{n}$, can be implemented as a unitary operator, in the reference Fock space, if and only if its antilinear part is  Hilbert-Schmidt \cite{Shale}, namely if and only if
\begin{align}\label{ucond2}
\sum_{n}g_{n}|\lambda_{n}^{f}(\eta)|^{2}<\infty \qquad \text{and} \qquad \sum_{n}g_{n}|\lambda_{n}^{g}(\eta)|^{2}<\infty,
\end{align}
for all times $\eta$. If this is true, then the two Fock representations defined for every value of the conformal time $\eta$ are unitarily related, and hence we can consider that the two analyzed quantizations are physically equivalent. 

It is straightforward to show that $\mathcal{K}_{n}(\eta)=\tilde{\mathcal{F}}_n(\eta)\mathcal{F}^{-1}_n(\eta)$, and that the norm of the off diagonal coefficients is then given by 
\begin{align}\label{kl}
|\lambda^{h}_{n}|=|\tilde{h}_{1}^{n}h_{2}^{n}-\tilde{h}_{2}^{n}h_{1}^{n}|.
\end{align}
We note that $|\lambda^{f}_{n}|=|\lambda^{g}_{n}|$, owing to relations \eqref{fgrel}. It is therefore sufficient to analyze the asymptotics of only one of these sequences of coefficients.

In addition, taking into account the symmetry provided by relations \eqref{fgrel}, it is general enough to just consider e.g. that the coefficients $\tilde{f}^n_l$ and $\tilde{f}^n_{\tilde l}$ are of the asymptotic form \eqref{hh} and \eqref{hh2}, respectively, where recall that $\{l,\tilde l\} = \{1, 2\}$ as a set. In the asymptotic limit of large $\omega_{n}$ we obtain that, for  $n\in\mathbb{N}^{+}_{l}$,
\begin{align}
|\lambda^{f}_{n}|=\frac{1}{\sqrt{2}}\left| \vartheta^{n}_{\tilde{f},l}+e^{i\tilde{F}^{n}_{l}}\Re(e^{-i\tilde{F}^{n}_{l}}\vartheta^{n}_{\tilde{f},l})\right| 
\end{align}
up to terms $\mathcal{O}(|\vartheta^{n}_{h,l}|^{2})$ if the field is massless, or up to terms $\mathcal{O}(\omega_{n}^{-1})$ if the field is massive and  $\{\sqrt{g_{n}}\omega_{n}^{-1}\}_{n\in\mathbb{N}^{+}_{l}}$ were square summable; otherwise, in the massive case, we have that
\begin{align}
|\lambda^{f}_{n}|=\frac{1}{\sqrt{2}}|\tilde \vartheta^{n}_{\tilde{f},l}|+\mathcal{O}(\omega_{n}^{-2}), \qquad n\in\mathbb{N}^{+}_{l}.
\end{align}
Since the sequences formed by $\vartheta^{n}_{\tilde{f},1}$ and $\vartheta^{n}_{\tilde{f},2}$ in the first case, and by $\tilde\vartheta^{n}_{\tilde{f},1}$ and $\tilde\vartheta^{n}_{\tilde{f},2}$ in the second case, are square summable in their respective subsets (including degeneracy), given our hypotheses on the unitary implementability of the quantum dynamics, then we see that conditions \eqref{ucond2} are immediately satisfied for all $n\in\mathbb{N}_{l}^{+}$.

On the other hand, it is not hard to check that $|\lambda^{f}_{n}|=\mathcal{O}(1)$ for all $n\in\mathbb{N}^{-}_{1}\cup\mathbb{N}^{-}_{2}$, owing to the difference in the relative sign of the coefficients in the pair $(\tilde{f}^{n}_{1},\tilde{f}^{n}_{2})$ with respect to that in $(f^{n}_{1},f^{n}_{2})$. So, in this case, the sequences that provide $\lambda^{f}_{n}$ and $\lambda^{g}_{n}$ are not square summable if any of the considered subsets of integers $\mathbb{N}^{-}_{l}$ has infinite cardinality. Then, the two analyzed Fock quantizations would not be unitarily equivalent. However, let us notice that an exchange in the roles of $\tilde f^n_l$ and $\tilde g^n_l$ would amount in practice to an interchange between the relative sign of the pair $(\tilde f^n_1,\tilde f^n_2)$ and the sign for the pair $(\tilde g^n_1,\tilde g^n_2)$. After this interchange, both pairs would display the same relative signs as the coefficients of our reference quantization [see Eq. \eqref{fgrel}]. But the discussed exchange of $\tilde f^n_l$ and $\tilde g^n_l$ can be regarded as a change in the convention of what are  particles and what antiparticles, as we see by inspecting Eq. \eqref{blockcs1}. Therefore, the inequivalence that we have found can be understood as artificially arising from the consideration of Fock quantizations with the opposite convention for the concept of particles and antiparticles in an infinite number of modes, with respect to our reference one. It is then clear that, once the conventions are reconciled, e.g. by switching the role of $f_{l}^{n}$ and $g_{l}^{n}$ for $n\in\mathbb{N}_{l}^{-}$ in the reference family of creation and annihilationlike variables, the physical phenomena described by the two considered quantizations would be equivalent. 

In summary, all Fock quantizations defined by invariant complex structures and related by a nontrivial dynamics that is unitarily implementable are either unitarily equivalent to the reference Fock quantization introduced above, or to one differing from it just in the convention adopted for the notion of particles and antiparticles in an infinite number of modes. In other words, the criterion of invariance under the symmetries of the equations of motion and the unitary implementability of a nontrivial quantum dynamics selects a unique class of unitarily equivalent Fock representations of the canonical anticommutation relations, together with a notion of quantum evolution, both for the massive and for the massless Dirac fields in 2+1 dimensions, up to conventions in the concepts of particles and antiparticles.

\section{Conclusions and discussion}
\label{sec:con}

In this paper we have extended previous results on the uniqueness of the Fock quantization of free Dirac fields propagating in certain nonstationary Lorentzian geometries in 3+1 dimensions with compact, connected, and orientable spatial sections \cite{uf}, to the lower dimensional case of 2+1 dimensions. We have shown that the criterion of invariance of the Fock vacuum under the symmetries of the equations of motion, together with the unitary implementability of a nontrivial dynamics, selects a unique class of unitarily equivalent Fock representations, both for the massive and for the massless fields, once the convention of particles and antiparticles is fixed (up to a finite number of modes at most). Furthermore, the criterion selects as well a quantum evolution for the fermion field, up to unitary redefinitions that are irrelevant in the ultraviolet sector, splitting in this manner the time variation of the Dirac field in two parts: a part consisting in an explicit time dependence, via a dependence on the background, and a part that can be promoted to a nontrivial quantum transformation. This last part, being unitarily implementable, maintains the coherence of pure states. In this sense, the time variation of the Dirac field can be separated in an essentially unique form into a background dependence, that does not respect coherence, and a quantum dynamics that is implementable as a unitary operator.

In order to carry out our study, we have started by expanding the fermion field in the basis of eigenspinors of the Dirac operator defined on the spatial sections. In that expansion, given in Eq. \eqref{expan}, we have introduced a scaling by the inverse of the scale-factor, so that the modes in the expansion, with coefficients denoted by $s_{np}$ and $\bar{r}_{np}$, are those associated with a conformally coupled field in absence of mass, $e^\alpha \Psi$, instead of the original minimally coupled field $\Psi$. These modes $s_{np}$ and $\bar{r}_{np}$ present constant canonical anticommutation relations with their complex conjugates. Then, we have fully analyzed the asymptotic regime of the fermion dynamics in the limit of large eigenvalues (in norm) of the Dirac operator.

We have analyzed the ambiguity inherent to the Fock quantization that is present, on the one hand, in the choice of a Fock representation of the canonical anticommutation relations and, on the other hand, in the splitting of the time dependence of the field in an explicitly time-varying part and a quantum evolution. It is this evolution what provides a specific family of time-dependent creation and annihilationlike variables that are dynamically related in the quantum theory. Different choices of complex structures and of a family of creation and annihilationlike variables define different Fock quantizations, that in general are unitarily inequivalent. We have then focused on those with vacua that are invariant under the unitary operators that implement the symmetries of the equations of motion of the fermion modes. These invariant vacua are characterized by creation and annihilationlike variables defined by $2\times 2$ matrices, that only depend on the number $n\in\mathbb{N}$ that labels eigenspaces of the Dirac operator with the same norm of the eigenvalue, $\omega_n$, and can only mix the modes $(s_{np}, \bar{r}_{np})$ (or their complex conjugates) with the same value of the degeneracy label $p$. We have characterized them by four coefficients, $f_l^n$ and $g_l^n$ with $l=1,2$, such that they satisfy the constraints \eqref{fgrel}.

Once we have established the form of the invariant creation and annihilationlike variables, we have determined the necessary and sufficient conditions that a time-dependent family of them has to verify in order that the variables at different times can be related by nontrivial dynamics that admit a unitary implementation. These conditions impose a very specific behavior on the coefficients $f_l^n$ and $g_l^n$ in the ultraviolet regime of large norm $\omega_n$ of the eigenvalue of the Dirac operator. In particular, in this asymptotic limit with respect to $\omega_n$, the leading order is completely fixed up to phase, with constant norm equal to $1/\sqrt{2}$. Depending on the spectral properties of the Dirac operator at hand, an additional subleading term of order $\omega_n^{-1}$ may be needed in the massive case. If this is so, it does not vanish only for two of the four coefficients, and its norm must equal $me^\alpha/(\sqrt{2}\omega_n)$. This term thus depends in a very specific way on the eigenvalue of the Dirac operator, on the mass of the field, and, moreover, on the dynamics of the spacetime background. The rest of terms, negligible compared to $\omega_n^{-1}$, are only constrained by the fact that they have to provide square summable sequences (including degeneracy). For a massless Dirac field, the term of order $\omega_n^{-1}$ is not needed, and therefore the unitarity requirement imposes only the conformal scaling of the Dirac field that is required to arrive at constant canonical Dirac brackets.

The equations of motion for the fermion modes might suggest that, in the ultraviolet regime, the problem reduces to that of the massless fermion field. As we have commented, this case displays conformal invariance, and admits a {\it natural}  quantization associated with the conformal vacuum. Therefore, and similarly to what happens for the scalar field case \cite{compact,compact2,unit}, at first glance one might think that, in the ultraviolet, conformal invariance is recovered, and that the conformal vacuum leads to a valid quantization in the massive scenario as well. However, this is not true owing to our requirement of getting a unitary dynamics for the massive Dirac field. Indeed, as discussed in the previous paragraph, the term of order $\omega_n^{-1}$, if present, turns out to be proportional to the mass of the field and cannot be neglected by any means in the ultraviolet regime. In 3+1 dimensions, we obtained the same result \cite{uf}, but there it is perhaps more clear since the beginning that the mass cannot be discarded in the ultraviolet regime, because it is the responsible of mixing the two chiralities that the Dirac field has in four dimensions. It is worth mentioning that there exist more similarities between the two dimensionally different problems. Indeed, it is as well true in the $3+1$ scenario (analyzed in Ref. \cite{uf}) that a unitarily implementable dynamics is achieved for the massless field by removing the background-dependent term of order $\omega_{n}^{-1}$ from the coefficients that characterize the allowed Fock quantizations of the massive case. Actually, the truncation at order one in one of the coefficients that define the resulting creation and annihilationlike variables amounts to the adoption of the conformal vacuum. The result of uniqueness of such a quantization follows almost straightforwardly.

We have finished our study by proving that two invariant Fock quantizations of the Dirac field with unitarily implementable dynamics indeed define quantum theories that are unitarily equivalent, up to conventions for the notions of particles and antiparticles. 

In conclusion, our criterion of symmetry invariance and unitarity of the quantum evolution is enough to remove the most serious ambiguities in the Fock quantization of Dirac fields propagating in the considered nonstationary geometries, with two-dimensional compact, connected, and orientable sections. On the other hand, comparison of our analysis with that of a fermion field in a spacetime of higher dimension, or with the case of a scalar field, provides a better understanding of the similarities and differences between these distinct situations. Finally, we hope that our work proves relevant in the context of fermion excitations in condensed matter physics, and in particular in the description of excitations in graphene \cite{graphene,gap}. We leave for the future the discussion of possible effective nonstationary geometries for those excitations, and the suggestion of tentative applications of our results to that field of research.

\acknowledgments
The authors are grateful to J. Gonz\'alez for helpful conversations. This work was partially supported by the research Grants No. MINECO Project No. FIS2014-54800-C2-2-P from Spain, No. DGAPA-UNAM IN113115 and No. CONACyT 237351 from Mexico, and COST Action No. MP1405 QSPACE, supported by COST (European Cooperation in Science and Technology). In addition, M. M-B acknowledges financial support from the Netherlands Organization for Scientific
Research (NWO), and from the Portuguese Foundation for Science and
Technology (FCT, Grant No. IF/00431/2015).

\appendix
\section{Asymptotics of the dynamics of the modes}
\label{app}

The dynamics of the modes $\{z_{np}\}=\{s_{np},r_{np}\}$ is governed by Eq. \eqref{2order}. In order to analyze their asymptotics, let us first look for a change of variables $z_{np}=f_n(\eta)\tilde{z}_{np}$ so that the equation of motion for the new variable $\tilde{z}_{np}$ does not involve its first derivative. It is straightforward to check that this is attained by choosing
\begin{align}
f_n(\eta)=\sqrt{\tilde{\omega}_n(\eta)}, \qquad \tilde\omega_n(\eta)=\omega_n+ime^{\alpha(\eta)}.
\end{align}
Then we obtain
\begin{align}\label{appeneq}
\tilde{z}_{np}^{\prime\prime}+\left[|\tilde{\omega}_{n}|^{2}-2\bigg(\frac{f_{n}'}{f_{n}}\bigg)^{2}+\frac{f_n^{\prime\prime}}{f_n}\right]\tilde{z}_{np}=0.
\end{align}
Let us search for two independent solutions $\tilde{z}_{np}^{l}$, with $l=1,2$, written in the form
\begin{align}\label{applaurent}
\tilde{z}_{np}^{l}=\exp{\left[-i(-1)^{l}\widetilde{\Theta}_{n}^{l}\right]}, \qquad \text{with} \qquad (\widetilde{\Theta}_{n}^{l})^{\prime}={\omega}_{n}+\Lambda_{n}^{l},
\end{align}
where $\Lambda_{n}^{l}$ is an $\omega_{n}$-dependent function of time that captures the freedom available to satisfy the dynamical equation \eqref{appeneq}. This equation of motion reduces to a  first-order differential equation of Riccati type
for the function $\Lambda_{n}^{l}$:
\begin{align}\label{lambdaeq}
(\Lambda_{n}^{l})^{\prime}=i(-1)^{l}\left[(\Lambda_{n}^{l})^{2}+2\omega_n\Lambda_{n}^{l}-e^{2\alpha}m^{2}+2\bigg(\frac{f_{n}'}{f_{n}}\bigg)^{2}- \frac{f_{n}^{\prime\prime}}{f_{n}}\right].
\end{align}

Let us now assume that $\Lambda^l_n=\mathcal{O}(\omega_n^{-1})$ in the ultraviolet limit, and that therefore, at leading order, it tends asymptotically to the solution $\tilde{\Lambda}_{n}^{l}$ of the equation
\begin{align}\label{lambdaeqb}
(\tilde\Lambda_{n}^{l})^{\prime}=i(-1)^{l}[2\omega_n\tilde\Lambda_{n}^{l}-e^{2\alpha}m^{2}]+ (-1)^{l} \frac{me^{\alpha}}{2\omega_{n}}[\alpha^{\prime\prime}+(\alpha^{\prime})^2].
\end{align}
In agreement with our previous comments (see footnote \ref{footnote3}), we assume that the second derivative of the scale factor, namely $e^{\alpha}[\alpha^{\prime\prime}+(\alpha^{\prime})^2]$, exists and is integrable in any closed interval of conformal time $[\eta,\eta_0]$.

An equation of this sort was analyzed in Ref. \cite{uf} (see also Ref. \cite{threeten}), excluding the last $\omega_{n}$-dependent inhomogeneous term that, with our conditions on the smoothness of the scale factor, and therefore on $\alpha$, can be seen to give a subdominant correction to the whole solution. Using the same line of reasoning as in that work, one can then show that the solution to Eq. \eqref{lambdaeqb} with initial condition $\tilde{\Lambda}_{n}^{l}(\eta_{0})=0$ is of order $\omega_{n}^{-1}$. Specifically, after an integration by parts, one obtains
\begin{eqnarray}\label{Lambdatilde}
\tilde{\Lambda}_{n}^{l}&=&\frac{m^{2}}{2\omega_{n}}\left(e^{2\alpha}-e^{2[\alpha_{0}+i(-1)^{l}\omega_{n}\Delta\eta]}\right)\nonumber\\
&-&\frac{m^{2}}{\omega_{n}}e^{2i(-1)^{l}\omega_{n}\eta}\int_{\eta_{0}}^{\eta}d\tilde{\eta}\alpha^{\prime}(\tilde{\eta})e^{2[\alpha(\tilde{\eta})-i(-1)^{l}\omega_{n}\tilde\eta]}+\mathcal{O}(\omega^{-2}_{n}).
\end{eqnarray}
We use the notation introduced before: $\Delta\eta=\eta-\eta_0$ and $\alpha_0=\alpha(\eta_0)$. Therefore, recalling our conditions on the behavior of $\alpha$, we conclude that there exists a positive function $C(\eta)$ which is $\omega_{n}$-independent and such that the absolute value of $\tilde{\Lambda}_{n}^{l}(\eta)$ is bounded by $C(\eta)/\omega_{n}$. Thus, the function $\tilde{\Lambda}_{n}^{l}$ is $\mathcal{O}(\omega^{-1}_{n})$, and can be taken as an asymptotic solution of Eq. \eqref{lambdaeq} up to higher-order corrections, consistently with our previous assumptions. 

The definition $ z_{np}=f_n \tilde z_{np}$ leads to the following relation between the function $\widetilde\Theta ^l_n$ introduced in Eq. \eqref{applaurent}, and the logarithm of the original solution, $\Theta ^l_n$, introduced in Eq. \eqref{sol}:
\begin{align}
(\Theta^{l}_{n})^{\prime}=\omega_n-\frac{(-1)^l m \alpha^{\prime}e^\alpha }{2(\omega_n+i m e^{\alpha})}+\Lambda^l_n.
\end{align}
With the initial condition $\Theta_{n,0}^{l}=\Theta_{n}^{l}(\eta_{0})=0$, we then get
\begin{align}
\Theta^{l}_{n}=\omega_{n}\Delta\eta + \frac{i}{2} (-1)^l \ln{\left( \frac{\omega_n+i m e^{\alpha}}{\omega_n+i m e^{\alpha_0}}\right)} +\int_{\eta_{0}}^{\eta}\text{d}\tilde\eta\Lambda^{l}_{n}(\tilde\eta).
\end{align}
On the other hand, with the initial condition $\Lambda_{n}^{l}(\eta_0)=0$, we obtain for $(\Theta^{l}_{n})^{\prime}$ the initial value
\begin{align}
\Theta^{l}_{n,1}=\omega_n-\frac{(-1)^l m \alpha^{\prime}_0e^{\alpha_0} }{2(\omega_n+i m e^{\alpha_0})},
\end{align} 
where $\alpha^{\prime}_0=\alpha^{\prime}(\eta_0)$. This in turn provides the following values for the quantities defined in Eq. \eqref{icconsts}:
\begin{align}
\Delta^{l}_{n}=\frac12-(-1)^l \frac{m \alpha^{\prime}_0e^{\alpha_0}}{4\omega_n(\omega_n+i m e^{\alpha_0})},\qquad \Gamma^{l}_{n}=\frac12+i\frac{m e^{\alpha_0}}{2\omega_n}.
\end{align}

\section{Asymptotic behavior of $\bar\Sigma^1_n-\Sigma^2_n$}
\label{appb}

We finally give here the asymptotic behavior of the difference $\bar\Sigma^1_n-\Sigma^2_n$, employed in Sec. \ref{sec:Unitarity} to analyze the beta coefficients. Recalling that $\Sigma^l_n=(\Theta^{l}_{n})^{\prime}-\omega_n$, we have from the results of Appendix \ref{app} that 
\begin{align}
\bar\Sigma^1_n-\Sigma^2_n=\bar\Lambda^{1}_n-\Lambda^{2}_n+\frac{\omega_n m \alpha^{\prime}e^\alpha }{\omega_n^2+m^2 e^{2\alpha}}.
\end{align}

Besides, we have seen that $\Lambda^l_n=\tilde\Lambda^{l}_n+c^l_n$ with $c^l_n=o(\omega_n^{-1})$. Actually, it is possible to show that $c^l_n= \mathcal{O}(\omega_n^{-2})$. From the differential equations \eqref{lambdaeq} and \eqref{lambdaeqb}, we obtain the following equation for $c^l_n$:
\begin{align}\label{c}
(c_{n}^{l})'=i(-1)^{l}\left\lbrace(c_{n}^{l})^{2}+2(\omega_n+\tilde\Lambda_{n}^{l})c_n^l + (\tilde\Lambda_{n}^{l})^2 -[2\alpha''\tilde{\omega}_{n}+(\alpha')^{2}(2\tilde{\omega}_{n}+3\omega_{n})]\frac{m^{2}e^{2\alpha}}{4\tilde{\omega}_{n}^{2}\omega_{n}}\right\rbrace,
\end{align}
where we recall that $\tilde{\omega}_n=\omega_n  + i m e^{\alpha}$. Since $c^l_n=o(\omega_n^{-1})$ and the inhomogeneous term of this equation is asymptotically $\mathcal{O}(\omega_n^{-2})$, it can be checked that the solution with initial condition $c_{n}^{l}(\eta_0)=0$ is indeed $\mathcal{O}(\omega_{n}^{-2})$ in the limit of large $\omega_{n}$. Noticing that, from Eq. \eqref{Lambdatilde}, it turns out that $\overline{\tilde\Lambda}\,^{1}_n-\tilde\Lambda^{2}_n=\mathcal{O}(\omega_{n}^{-2})$, we then arrive at
\begin{align}
\bar\Sigma^1_n-\Sigma^2_n=\frac{m\alpha^{\prime} e^\alpha }{\omega_n}+\mathcal{O}(\omega_n^{-2}).
\end{align}
Obviously, its integral is the quantity given in Eq. \eqref{sig}.

\end{document}